\documentclass[reprint,aps,prl,twocolumn,superscriptaddress]{revtex4-1}
\usepackage{braket}
\usepackage{amssymb}
\usepackage{amsmath}
\usepackage{epsfig}
\usepackage{color}
\usepackage{graphics, graphicx}
\usepackage{bbold}
\usepackage{psfrag}
\usepackage{mathcomp}
\usepackage{subfigure}
\usepackage{verbatim}
\usepackage[colorlinks, citecolor=blue]{hyperref}
\usepackage[normalem]{ulem}
\usepackage[compat=1.1.0]{tikz-feynman}

\def\cp#1{\mathbf{#1}}
\begin{document}

\title{Liquid-Gas Transition and Coexistence in Ground State Bosons with Spin Twist}
\author{Qi Gu}
\email{qigu@tsinghua.edu.cn}
\affiliation{Institute for Advanced Study, Tsinghua University, Beijing 100084, China}
\author{Xiaoling Cui}
\email{xlcui@iphy.ac.cn}
\affiliation{Beijing National Laboratory for Condensed Matter Physics, Institute of Physics, Chinese Academy of Sciences, Beijing 100190, China}

\date{\today}

\begin{abstract}
We study the thermodynamic liquid-gas transition and coexistence (LGTC) for ground state bosons under contact interactions. We find that the LGTC can be  facilitated by the mismatch of spin polarization, dubbed ``spin twist", between single-particle and interaction channels of  bosons with spin degree of freedom. Such spin twist uniquely stabilizes the gas phase by creating an effective repulsion for low-density bosons,  thereby enabling LGTC in the presence of a quantum droplet at much larger density.  
We have demonstrated the scheme for binary bosons subject to Rabi coupling and  magnetic detuning, where the liquid-gas transition can be conveniently tuned and their coexistence can be characterized by  discontinuous density profile in a harmonic trap. The spin twist scheme for LGTC can be generalized to a wide class of quantum systems with competing single-particle and interaction orders.
\end{abstract}
\maketitle

{\it Introduction.}
Liquid-gas transition and coexistence (LGTC)\cite{Chaikin1995:PrinciplesCondensedMatter} appear to be a common physical phenomenon in nature. Nowadays the phenomenon has got many industrial applications in oil, natural gas, aerospace, chemical engineering etc, and its research has recently even extended to hot nuclei\cite{nuclear1, nuclear2, nuclear3} and active matters\cite{active_matter1,active_matter2}. All these systems have two common features. Namely, the LGTCs therein all occur at finite temperature($T$) within certain $T$-window, and they are all associated with long range interactions characterized by a repulsive core and an attractive tail (e.g., Lennard-Jones potential), as responsible for the liquid stabilization. Indeed, a textbook model for LGTC is based on the Van der Waals' equation of state\cite{Landau_Lifshitz, Huang1991:StatisticalMechanics2nd}, exactly reflecting the important roles played by thermal effect and long-range potential. 

The question is can we go beyond the traditional frame to engineer LGTC, so as to broaden the understanding on its nature and enable its potential use in different systems. As a first attempt, Miller {\it et al} showed that the liquid-gas transition can occur at zero $T$ where the quantum statistics played a vital role\cite{zero_T_1, zero_T_2}. Nevertheless, they still relied on the long-range interaction and concluded the absence of liquid-gas coexistence in bosons given the associated transition is of second order. On the other hand, the recent realization of quantum droplet in ultracold gases\cite{Ferrier-Barbut2016:ObservationQuantumDroplets,Schmitt2016:SelfboundDropletsDilute,Chomaz2016:QuantumFluctuationDrivenCrossoverDilute,Cabrera2018:QuantumLiquidDroplets,Cheiney2018:BrightSolitonQuantum,Semeghini2018:SelfBoundQuantumDroplets,DErrico2019:ObservationQuantumDroplets, Modugno,Pfau_4,Ferlaino_2, DJWang} offers an unprecedented opportunity for addressing the question.   These ultracold droplets (resemblance of liquids) are stabilized by an attractive mean-field interaction and a repulsive force from quantum fluctuations\cite{Petrov2015:QuantumMechanicalStabilization}, where  the interaction  is not necessarily long range but can be a contact one. 
However, the phenomenon of LGTC has not been deterministically observed in these systems up to date. It is essentially because the liquid-gas transition therein  is continuous at zero $T$\cite{footnote}, which cannot host any coexistence region as in accordance with Refs.\cite{zero_T_1, zero_T_2}. Note that the  transitions measured in existing experiments\cite{Cabrera2018:QuantumLiquidDroplets,Cheiney2018:BrightSolitonQuantum,Semeghini2018:SelfBoundQuantumDroplets,DErrico2019:ObservationQuantumDroplets, DJWang} are driven by the quantum pressure of finite-size system, instead of a thermodynamic one. Moreover, since the quantum droplet is quite fragile to thermal effect \cite{Ota2020:LeeHuangYangDescriptionSelfbound,Wang2021:UltradiluteSelfboundQuantum,Guebli2021:QuantumSelfboundDroplets}, the LGTC at finite $T$, even exists, can be quite difficult to detect given the expected narrow $T$-window in reality.

\begin{figure}[!ht]
    \centering
   \includegraphics[width=8.5cm]{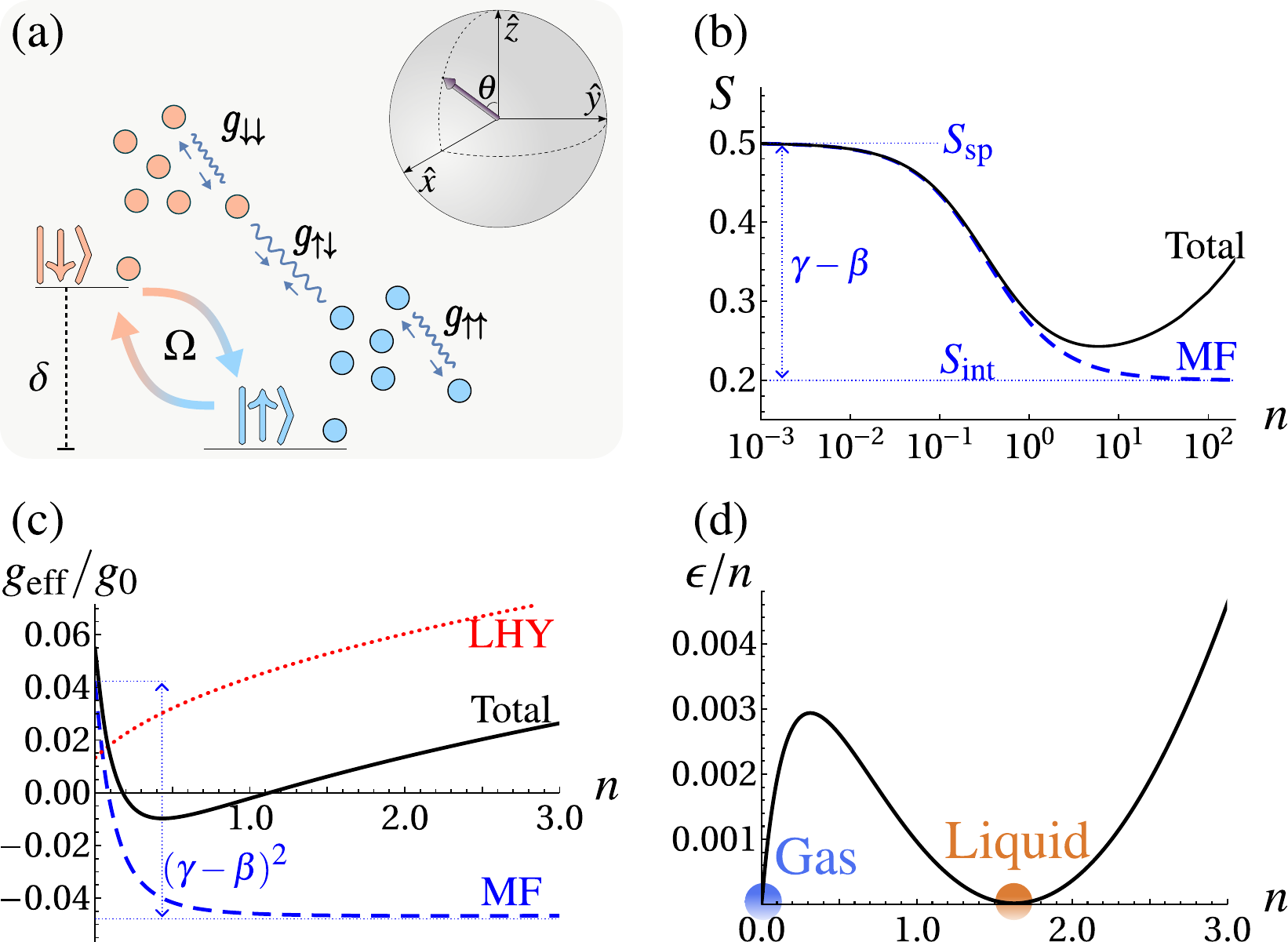}
    \caption{(a) Schematics of a spin twist setup. The spin state of binary bosons, $(\psi_{\uparrow},\psi_{\downarrow})\propto (\cos(\theta/2),\sin(\theta/2))$, can be mapped onto a Bloch sphere with polarization $S\equiv\langle \sigma_z\rangle = \cos\theta$. Single-particle potentials $\{\Omega,\ \delta\}$ and spin-dependent interactions $\{g_{\sigma\sigma'}\}$ respectively optimize $\theta$ as $\theta_{\rm sp}$ and $\theta_{\rm int}$. A 'spin twist' occurs when $\theta_{\rm sp}\neq\theta_{\rm int}$ and thus $S_{\rm sp}\neq S_{\rm int}$. (b) Density-tuned polarization from $S_{\rm sp}(=\gamma)$ to  $S_{\rm int}(=\beta)$.  Dashed and solid lines respectively show mean-field and total (with LHY correction) results. (c) Effective interaction (black solid line) and its individual contribution from mean-field(blue dashed) and LHY(red dot) sectors. The spin twist leads to an additional mean-field repulsion in low-density regime ($\sim(\gamma-\beta)^2$), as marked by blue vertical line, which uniquely stabilizes the gas state.  (d) Energy per particle (shifted by single-particle energy $-\sqrt{\Omega^2+\delta^2}$) as a function of density, where the double minima  indicate liquid-gas coexistence near their first-order transition.  
In (b,c,d) we take parameters $(\alpha, \beta,\gamma,\eta)=(-0.1, 0.2, 0.5, 0.0137)$, and scale the density and energy per particle respectively by $\Omega/g_0$ and $\Omega$. }\label{fig_twist}
\end{figure}

In this work, we unveil a new mechanism for the LGTC of ground state bosons (zero $T$) with contact interaction, thereby well beyond the traditional frame as well as previous theories\cite{zero_T_1, zero_T_2}. 
Such mechanism is based on ``spin twist", which refers to  a mismatch of spin polarization between single-particle and interaction channels for bosons with spin degrees of freedom. 
To demonstrate the idea, we consider a concrete setup of binary (pseudo-spin $1/2$) ultracold bosons subject to  Rabi coupling ($\Omega$) and magnetic detuning($\delta$), see schematics in Fig.\ref{fig_twist}(a), as explored in previous experiments\cite{Lavoine2021:BeyondMeanFieldEffectsRabiCoupled, Hammond2022:TunableThreeBodyInteractions,Oberthaler2011,Tarruell2022,Ferrari}. The single-particle potentials ($\Omega,\ \delta$) determine an optimal spin polarization, which can be tuned to mismatch the one determined by spin-dependent interactions (Fig.\ref{fig_twist}(b)). Such spin twist
leads to an effective repulsion for low-density bosons (Fig.\ref{fig_twist}(c)), which uniquely stabilizes the gas phase and renders the first-order LGTC in the presence of a quantum droplet at much larger density (Fig.\ref{fig_twist}(d)). In this case, the resulted liquid-gas transition can be conveniently tuned by $\Omega,\ \delta$ and interaction strengths, and moreover, they all occur for thermodynamic systems, in contrast to the finite-size transitions observed previously\cite{Cabrera2018:QuantumLiquidDroplets,Cheiney2018:BrightSolitonQuantum,Semeghini2018:SelfBoundQuantumDroplets,DErrico2019:ObservationQuantumDroplets, DJWang}. 
To characterize the liquid-gas coexistence, we have pointed out their phase separation in a harmonic trap 
and further identified two universal exponents for the critical scaling of their densities. Our results can be readily detected in ultracold experiments, and the spin twist scheme can serve as a general tool to engineer LGTC in quantum systems. 

{\it Model.}   
We consider the binary bosons ($\uparrow,\downarrow$) with Hamiltonian $H=H_0+U$: ($\hbar=1$)
\begin{eqnarray}
H_0&=&\int d{\cp r}\sum_{\sigma\sigma'}\psi_{\sigma}^\dagger({\cp r}) \left(-\frac{\nabla^2}{2m}\delta_{\sigma\sigma'} - [\Omega\sigma_x+\delta \sigma_z]_{\sigma\sigma'}\right) \psi_{\sigma'}({\cp r}) ;\nonumber\\
U&=&\frac{1}{2} \int d{\cp r} \sum_{\sigma\sigma'}g_{\sigma\sigma'}\psi_{\sigma}^\dagger({\cp r})\psi_{\sigma'}^\dagger ({\cp r})\psi_{\sigma'}({\cp r})\psi_{\sigma} ({\cp r}). \label{Hamiltonian}
\end{eqnarray}
Here $ \psi_{\sigma}^\dagger(\mathbf{r}) $ is the creation operator of spin-$\sigma$, and $\sigma_{i}$ ($i=x,y,z$) are Pauli matrices;  $\Omega$ and $\delta$ are respectively the strengths of Rabi coupling and magnetic detuning; $g_{\sigma\sigma'}$ is the contact coupling strength between $\sigma$ and $\sigma'$, and here we consider $g_{\uparrow\uparrow},g_{\downarrow\downarrow}>0$ and $\delta g\equiv g_{\uparrow\downarrow}+\sqrt{g_{\uparrow\uparrow}g_{\downarrow\downarrow}}<0$, where a stable quantum droplet can be supported in the absence of $\Omega$ and $\delta$\cite{Petrov2015:QuantumMechanicalStabilization}. The multiple parameters in this problem can be recombined into four dimensionless ones: 
\begin{eqnarray}
\alpha\equiv \frac{\delta g}{g_0}; \ \ \beta\equiv \frac{g_{\downarrow\downarrow}-g_{\uparrow\uparrow}}{4g_0}; \ \ \gamma\equiv \frac{\delta}{\sqrt{\delta^2+\Omega^2}};\ \ \eta\equiv m^3\Omega g_0^2;\nonumber
\end{eqnarray}
with $g_0\equiv (g_{\uparrow\uparrow}+g_{\downarrow\downarrow}-2g_{\uparrow\downarrow})/4$. Here $\alpha$ characterizes the strength of overall attractive interaction; 
$\beta$ and $\gamma$, as shown later, stand for the optimal spin polarizations in interaction and single-particle channels, respectively; $\eta$ measures the Rabi field with respect to interaction strength. To simplify the discussions, in this work we  mainly consider the effects of tunable $\alpha$ and $\gamma$. 

{\it Spin twist and the induced effective repulsion.}
Under the mean-field treatment, we replace the field operators by classical numbers: $\psi_{\uparrow}=\sqrt{n}\cos(\theta/2)$, $\psi_{\downarrow}=\sqrt{n}\sin(\theta/2)$, where $n$ is the total density, and $\theta$ determines the spin polarization 
\begin{equation}
S\equiv\frac{n_{\uparrow}-n_{\downarrow}}{n_{\uparrow}+n_{\downarrow}}=\cos\theta. \label{P}
\end{equation}  
The mean-field energy per volume, $\epsilon_{\rm mf}=E_{\rm mf}/V$, is 
\begin{equation}
\epsilon_{\rm mf}=-(\sqrt{1-S^2}\Omega+ S\delta)n + \frac{g_0n^2}{2}\left[(S-\beta)^2+\frac{g_{\uparrow\uparrow}g_{\downarrow\downarrow}-g_{\uparrow\downarrow}^2}{4g_0^2}\right]. \label{e_mf}
\end{equation}
Clearly, the first term contributed from single-particle potentials favors spin polarization $S_{\rm sp}=\gamma$, while the second term from interactions  favors  $S_{\rm int}=\beta$. A ``spin twist" occurs when the two polarizations mismatch, i.e., $\beta\neq\gamma$. 
The overall mean-field polarization, as determined by $\partial \epsilon_{\rm mf}/\partial S=0$, is shown by dashed line in Fig.\ref{fig_twist}(b), which is density-dependent and can change from $\gamma$ to $\beta$ as $n$ increases.

A remarkable effect of such spin twist is to induce an effective repulsion uniquely in low-density limit.  Here we define the effective interaction as
\begin{equation}
g_{\rm eff}\equiv \frac{\partial^2\epsilon}{\partial n^2},
\end{equation}
where $\epsilon$ is the energy density after optimizing $S$. In the absence of spin twist, the two terms in Eq.\ref{e_mf} both favor $S=\beta=\gamma$, leading to $g_{\rm eff,mf}^{(0)}=
(g_{\uparrow\uparrow}g_{\downarrow\downarrow}-g_{\uparrow\downarrow}^2)/(4g_0)$. This is the conventional case of binary bosons, whose mean-field stability is given by  $g_{\uparrow\uparrow}g_{\downarrow\downarrow}>g_{\uparrow\downarrow}^2$ for any density. However, it is no longer true when spin twist occurs ($\beta\neq \gamma$). In this case, the single-particle and interaction terms compete with each other and the resulted $S$ and $g_{\rm eff,mf}$ are generally $n$-dependent. In the low $n$ limit, the single-particle terms dominate, which result in $S\sim \gamma$ and 
\begin{equation}
g_{\rm eff,mf}=g_{\rm eff,mf}^{(0)}+{g_0}(\gamma-\beta)^2. \label{geff}
\end{equation}
Here the spin twist leads to an additional repulsion $\sim {g_0}(\gamma-\beta)^2$ at the mean-field level. Its physical origin can be understood as follows:  
since the interactions favor $S\sim \beta$ as ground state, here $S\sim \gamma$ corresponds to an excited spin orientation in the interaction channel, which naturally generate an effective repulsion as above. 
Note that such repulsion only works for low densities but not high ones, where the interactions dominate and recover $S\sim\beta$ and $g_{\rm eff,mf}\sim g_{\rm eff,mf}^{(0)}$.

Beyond the mean-field treatment, we have further carried out the Bogoliubov analysis and extracted the Lee-Huang-Yang (LHY) energy $\epsilon_{\rm LHY}$ from quantum fluctuations\cite{supple}. 
In Fig.\ref{fig_twist}(c), we plot out the typical $n$-dependent $g_{\rm eff}$ obtained from the total $\epsilon=\epsilon_{\rm mf}+\epsilon_{\rm LHY}$, as well as its individual contributions from mean-field and LHY parts.  
As expected, the mean-field contribution $g_{\rm eff, mf}$ is positive only in low $n$ limit, and gradually reduces to a  negative value as $n$ increases. The reduction is exactly given by $\sim g_0(\gamma-\beta)^2$  due to spin twist (Eq.\ref{geff}). 
In comparison, the LHY contribution $g_{\rm eff, LHY}$ is always positive and continuously grows with $n$. The total $g_{\rm eff}$ then displays intriguing sign-changing as $n$ increases:  it turns from positive to negative, and back to positive again at large $n$. Consequently, the energy per particle $\epsilon/n$ as a function of $n$ displays double minima, see Fig.\ref{fig_twist}(d). Here the second minimum  at finite $n$ is  a self-bound  droplet with zero pressure that is mainly balanced by mean-field attraction and LHY repulsion, sharing the same spirit as ordinary case \cite{Petrov2015:QuantumMechanicalStabilization}. While the first minimum at $n=0$, newly found in this work, stands for a gas stabilized by unique low-$n$ repulsion under spin twist (c.f. Eq.\ref{geff}). Such double minima structure  implies a first-order transition between liquid and gas as well as their coexistence under proper conditions. We have checked that similar structure cannot appear without spin twist\cite{supple}. 

From the dressed basis perspective, the gas with $n\sim 0$ and $S\sim \gamma$ essentially occupies the lowest dressed branch as explored in experiments\cite{Lavoine2021:BeyondMeanFieldEffectsRabiCoupled, Hammond2022:TunableThreeBodyInteractions, Tarruell2022}; while for the liquid phase at finite $n$, whose polarization can be far from $\gamma$, the single dressed branch is generally not a good description due to considerable population at higher branch. 



\begin{figure}[!ht]
    \centering
   \includegraphics[width=8.5cm]{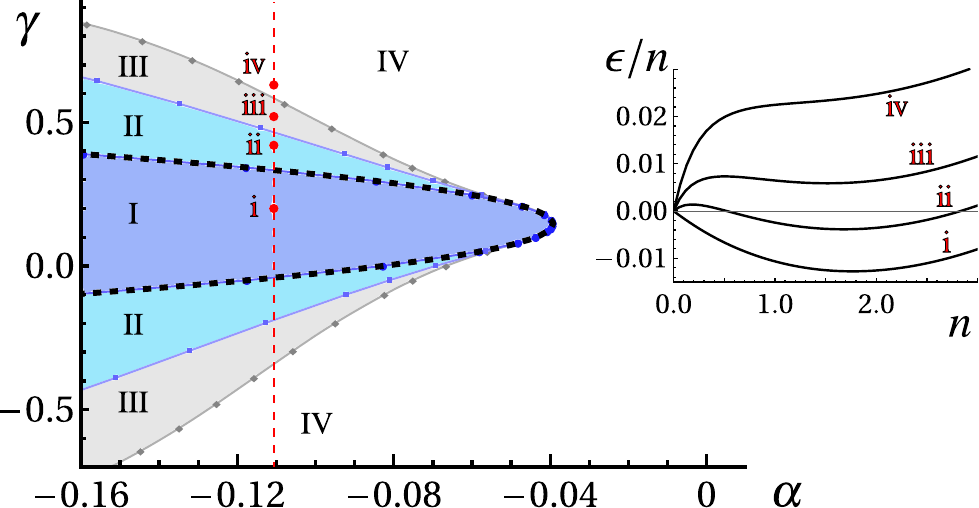}
   \caption{ Ground state phase diagram in ($\alpha, \gamma$)  plane with fixed $\beta=0.141, \eta=0.0157$.  Here I,II,III,IV respectively denote the region where the ground state is a pure droplet, a droplet with metastable gas,  a gas with metastable droplet, and a pure gas.  The liquid-gas transition occurs at II-III boundary. The inset shows typical $\epsilon/n (n)$ curves for different regions, as marked by 'i,ii,iii,iv' in the main plot. For each curve $\epsilon/n$ is shifted  by single-particle energy ($-\sqrt{\Omega^2+\delta^2}$). We scale $n$ and $\epsilon/n$ respectively by $\Omega/g_0$ and $\Omega$. }
    \label{fig_diagram}
\end{figure}

{\it Ground state phase diagram.} The ground state  (gas or self-bound droplet) is given by the global minimum in $\epsilon/n (n)$ curve, which is associated with the lowest chemical potential $\mu=\partial\epsilon/\partial n=\epsilon/n$. In Fig.\ref{fig_diagram}, we present the ground state phase diagram in $(\alpha, \gamma)$ plane for a given set of $\beta,\eta$. Four phases are shown, i.e.,  pure droplet (I), droplet with metastable gas (II), gas with metastable droplet (III), and pure gas (IV). Typical $\epsilon/n (n)$ landscapes for different regions are given in the inset plot. The double minima structure appears in regions II and III, and the  liquid-gas transition occurs at II-III boundary when they have the same $\epsilon/n=\mu=-\sqrt{\Omega^2+\delta^2}$, i.e., the single-particle shift. 

We would like to remark a crucial difference between the liquid-gas transition here and those observed previously in binary bosons\cite{Cabrera2018:QuantumLiquidDroplets,Cheiney2018:BrightSolitonQuantum,Semeghini2018:SelfBoundQuantumDroplets,DErrico2019:ObservationQuantumDroplets, DJWang}. In previous cases, the  transition is driven by the gradually dominant quantum pressure as compared to interaction terms when the boson number $N$ decays, and therefore it occurs for finite-size systems when $N$ reaches a critical value.  However, in our case the transition  occurs in the thermodynamic limit ($N,V\rightarrow\infty$ with $n=N/V$) and is driven by the competition between single-particle and interaction potentials. Therefore, the current case allows a highly tunable transition point for an arbitrarily large system, and moreover, allows the exploration of liquid-gas coexistence in a considerably broad parameter regime,  as shown below.

{\it Liquid-gas coexistence.} We now analyze the properties of bosons confined in a harmonic trap.  We consider a realistic system of $^{39}$K atoms with hyperfine states $|F=1,m_F=-1\rangle\equiv |\uparrow\rangle,\ |F=1,m_F=0\rangle\equiv |\downarrow\rangle$, as well studied in ultracold droplet experiments\citep{Cabrera2018:QuantumLiquidDroplets,Cheiney2018:BrightSolitonQuantum,Semeghini2018:SelfBoundQuantumDroplets}. In this system, $a_{\uparrow\uparrow}=35a_0,\ a_{\uparrow\downarrow}=-53a_0$ ($a_0$ is the Bohr radius), and $a_{\downarrow\downarrow}$ is highly tunable by magnetic field. For a concrete demonstration, here we take $a_{\downarrow\downarrow}=64a_0$, $\Omega=(2\pi)3.5$kHz (thus $\alpha, \beta, \eta$ are all fixed), and only focus on the coexistence region tuned by $\delta$ (or $\gamma$). 

\begin{figure}[!ht]
    \centering
    \includegraphics[width=8.5cm]{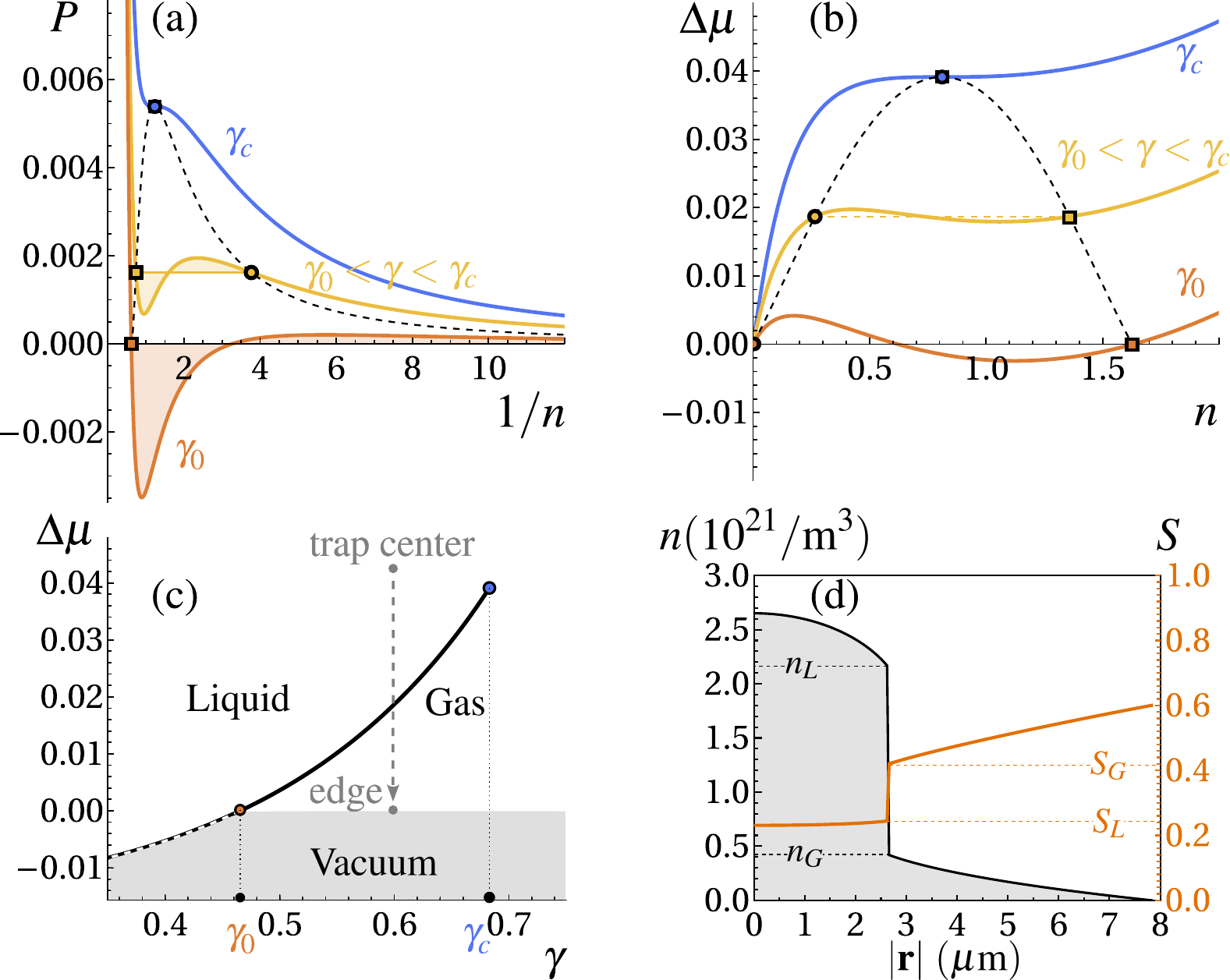}
\caption{Liquid-gas coexistence(LGC) tuned by $\gamma$ at fixed $\alpha=-0.11,\beta=0.141,\eta=0.0157$.  (a) Pressure $P$ as a function of $1/n$ for different $\gamma=0.465(=\gamma_0),\ 0.6,\ 0.683(=\gamma_c)$ (from bottom to top). (b) Shifted chemical potential $\Delta\mu\equiv \mu+\sqrt{\Omega^2+\delta^2}$ as a function of $n$ for different $\gamma$ as in (a). The intersections between these curves and horizontal lines in (a,b) give the equilibrium densities of liquid ($n_L$, squares) and gas ($n_G$, circles), which are connected by binodal lines (dashed).  For each curve in (a), the two shadow regions have the same area following the Maxwell's construction. (c) Phase diagram of liquid, gas and vacuum in the ($\gamma,\Delta \mu$) plane. LGC occurs along the black line 
for $\gamma\in(\gamma_0,\gamma_c)$. At $\gamma<\gamma_0$, only a self-bound droplet (liquid) is present if $\Delta\mu$ is above the dashed line. At $\gamma>\gamma_c$, the liquid and gas are indistinguishable. 
(d) Profiles of total density($n$) and spin polarization($S$) for bosons  in an isotropic harmonic trap with frequency $\omega=(2\pi)50$Hz, total number $N=4\times 10^5$ and $\gamma=0.6$, corresponding to the vertical trajectory shown in (c). The sharp jumps of $n$ and $S$ mark the location of LGC. For all plots, we scale the density and energy per particle respectively by $\Omega/g_0$ and $\Omega$.}
    \label{fig_LGC}
\end{figure}

The coexistence of liquid and gas requires 
\begin{equation}
\mu(n_L)=\mu(n_G),\ \ \ \ \ \ P(n_L)=P(n_G); \label{condition}
\end{equation}
where $n_L\ (n_G)$ is the liquid (gas) density at equilibrium, $\mu$ is the chemical potential and $P=\mu n-\epsilon$ is the pressure. 
In Fig.\ref{fig_LGC}(a,b), we plot out $P (1/n)$ and $\mu (n)$ for several typical $\gamma$. 
We can see that $P (1/n)$ shows  similar lineshape as classical $P$-$V$ isotherms hosting LGTC\cite{Landau_Lifshitz, Huang1991:StatisticalMechanics2nd}. Here we have used the Maxwell's construction to identify $n_L$ and $n_G$, as marked respectively by squares and circles in Fig.\ref{fig_LGC}(a). Specifically,  $n_L$  ($n_G$) is given by the left  (right) intersection between each $P (1/n)$ curve and a horizontal line, by requiring the same area of two separated shadow regions. 
The relation $\int d\mu=\int 1/n dP$ then guarantees the same $\mu$ at the intersections (see also Fig.\ref{fig_LGC}(b)), which will be denoted as $\mu_{\rm LGC}$ from now on.

Fig.\ref{fig_LGC}(a,b) also indicate a finite parameter window, $\gamma\in(\gamma_0,\gamma_c)$, for the occurrence of liquid-gas coexistence. At the lower bound $\gamma_0$, a gas phase starts to emerge at  $n_{G}=0$ and the two phases have $\mu=-\sqrt{\Omega^2+\delta^2}$  and $P=0$. This is right at the ground state transition between liquid and gas, i.e., at the II-III  boundary  shown in Fig.\ref{fig_diagram}. As increasing $\gamma$, the coexisting phases has higher $\mu,\ P$  and meanwhile $n_G$ and $n_L$ get closer. Finally the coexistence terminates at the critical point $\gamma_c$ when $n_L,\ n_G$ merge into a single value($n_c$) at the inflections of $\mu (n)$ and $P (1/n)$ curves. For even larger $\gamma>\gamma_c$, the gas and liquid become indistinguishable. 

Fig.\ref{fig_LGC}(c) further summarizes the results in  $(\gamma,\mu)$ plane, where $\mu_{\rm LGC}$ (solid line) separates the liquid and gas for $\gamma\in(\gamma_0,\gamma_c)$. 
To observe their coexistence, we suggest measuring the density profiles of bosons under an external trap, and here for brevity we consider an isotropic harmonic trap $V({\cp r})=m\omega^2{\bf r}^2/2$. 
Using the local density approximation $\mu({\cp r})=\mu(0)-V({\cp r})$, in Fig.\ref{fig_LGC}(d) we plot out the typical profiles of $n$ and $S$ in the trap showing liquid-gas phase separation, where liquid and gas respectively occupy the trap center and edge. At their interface  $n$ ($S$) displays sharp jump from $n_L$ to $n_G$ ($S_L$ to $S_G$), marking the location of liquid-gas coexistence with $\mu=\mu_{\rm LGC}$. Note that $n_{L,G}$ and $S_{L,G}$ do not depend on specific boson number $N$, as long as it is above a critical value\cite{supple}.

\begin{figure}[!ht]
    \centering
    \includegraphics[width=8.5cm]{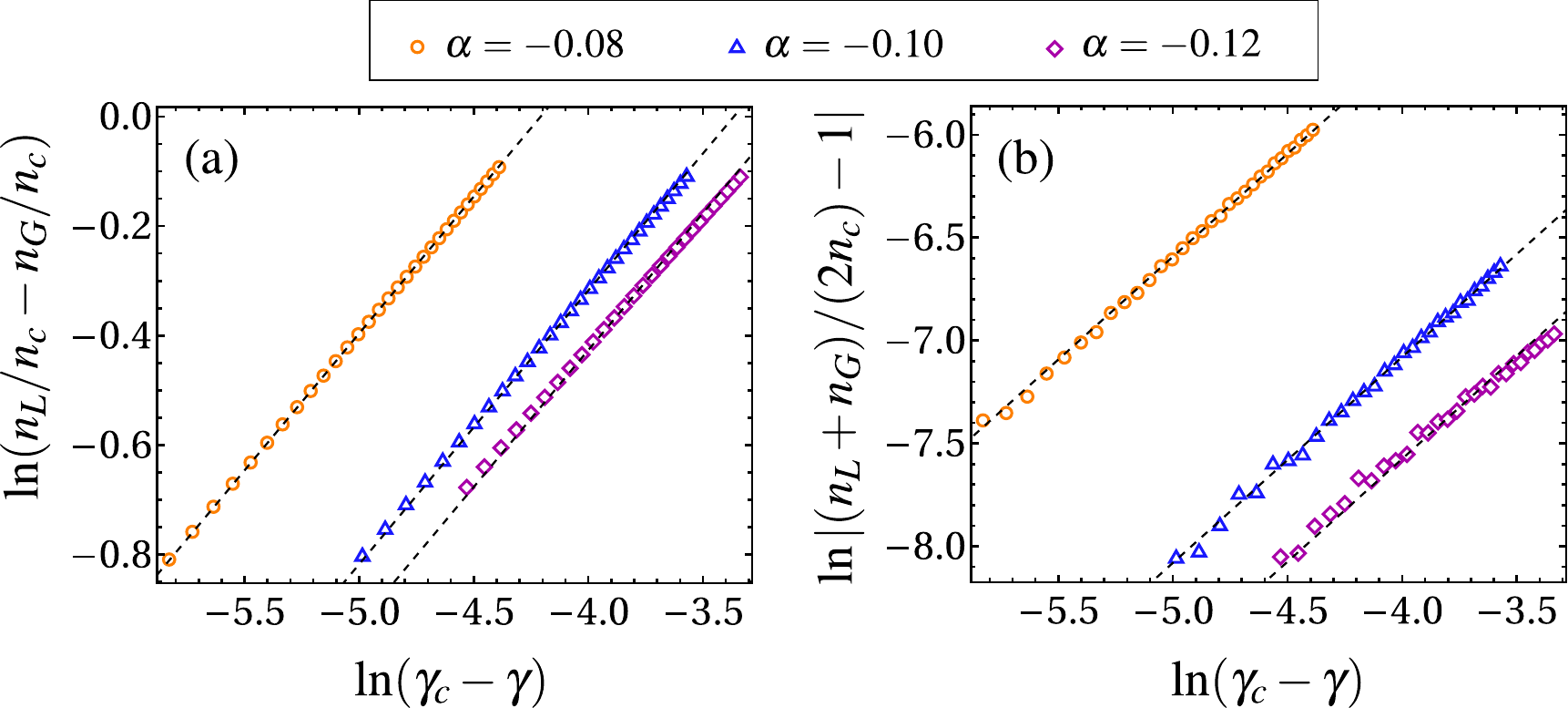}
    \caption{Universal critical scaling for the relative (a) and averaged (b) densities of liquid and gas at their coexistence. $\beta,\eta$ are the same as in Fig.\ref{fig_LGC}. Discrete points show numerical data and dashed lines show linear fittings. 
    Here we take $\ln$-$\ln$ plot, and the slopes for all fitting lines in (a) are $1/2$ and in (b) are $1$, giving the according exponents $\lambda$ and $\xi$ defined in Eq.(\ref{scaling}).      }
    \label{fig_scaling}
\end{figure}

Interestingly, the liquid and gas obey universal critical scaling near the melting of LGTC. Here we explore the asymptotic behavior of their relative and mean densities near $\gamma\sim \gamma_c$ and $n_L\sim n_G\sim n_c$:
\begin{equation}
\frac{n_L-n_G}{n_c}\propto (\gamma_c-\gamma)^{\lambda}; \ \ \ \frac{n_L+n_G}{2n_c}-1 \propto (\gamma_c-\gamma)^{\xi}, \label{scaling}
\end{equation}
with $\lambda,\ \xi$ the according critical exponents.  In Fig.\ref{fig_scaling}(a,b), we have numerically extracted the exponents as $\lambda=1/2$ and $\xi=1$ for all given $\alpha$. These exponents also universally persist for other tunable parameters, 
such as changing $\gamma$ to $\alpha$\cite{supple}. To explain such universal phenomenon, we have adopted a mean-field theory as in the classical treatment of finite $T$ liquid-gas transition\cite{Chaikin1995:PrinciplesCondensedMatter}, which well 
predicts the universal critical exponents  as above\cite{supple}.   

Here we clarify that the liquid-gas coexistence here should be distinguished from the equilibrium of a  droplet and a fully polarized gas in previous studies\cite{LGC1,LGC2,LGC3, LGC4}, where $\Omega,\delta$ are both absent. The latter is due to a preset spin population that deviates from the one preferred by droplet, and therefore the residue (single-species) bosons are repelled out of the droplet to form gas. In contrast, the coexistence here is associated with first-order liquid-gas transition (Fig.\ref{fig_twist}), where the spin population is changeable and both species can transfer freely between two phases to reach mutual equilibration.  

{\it Summary.} We have revealed a new mechanism using spin twist to engineer liquid-gas transition and coexistence (LGTC) in ground state bosons, which does not rely on thermal effect  or long-range potential.  The scheme is demonstrated with a specific model of binary bosons under Rabi coupling and magnetic detuning. Take the realistic $^{39}$K atoms for example, in practice one can follow the  strategy in \cite{Lavoine2021:BeyondMeanFieldEffectsRabiCoupled, Hammond2022:TunableThreeBodyInteractions, Tarruell2022} to prepare a gas  at ground state, from which a liquid (droplet) can be approached by changing $\gamma$ or $\alpha$ following Fig.\ref{fig_diagram}. The LGTC naturally occur during this process, and the proposed  phase separation in a trap and universal scaling of equilibrium densities can be readily tested in experiments.

Finally, we remark that the spin twist in creating an effective low-density repulsion for gas stabilization is a very robust mechanism, which  can be applied to a wide class of quantum systems with competing single-particle and interaction orders. For instance, it is expected to still work when add more spin degrees of freedom, or  change $\Omega,\delta$ to other single-particle potentials in altering spins. In this regard,  the spin twist can serve as a general principle for achieving LGTC at ultra-low temperatures, which hopefully would promote the practical use of such phenomena in a fascinating quantum world in future.

\acknowledgments
{\bf Acknowledgment.}
The work is supported by the National Key Research and Development Program of China (2018YFA0307600), the National Natural Science Foundation of China (12074419, 12134015), and the Strategic Priority Research Program of Chinese Academy of Sciences (XDB33000000).

\clearpage
\onecolumngrid
 \vspace*{1cm}
\begin{center}
{\large\bfseries Supplementary Materials}
\end{center}
\setcounter{figure}{0}
\setcounter{equation}{0}
\renewcommand{\figurename}{Fig.}
\renewcommand{\thefigure}{S\arabic{figure}}
\renewcommand{\theequation}{S\arabic{equation}}

This supplemental file includes the derivations of beyond-mean-field correction, details on liquid-gas coexistence in trapped system, and theoretical explanation of universal scaling near the critical point.  

\section{Beyond-mean-field correction}
\subsection{The Bogoliubov method and LHY correction}
The Hamiltonian in the momentum space is $H=H_0+U$, where
\begin{equation}\label{Hamiltonian-k-space}
\begin{aligned}
	H_0 &= \sum_\mathbf{k}\left[\epsilon^0_\mathbf{k}(\psi^\dagger_{\uparrow,\mathbf{k}}\psi^{}_{\uparrow,\mathbf{k}}+\psi^\dagger_{\downarrow,\mathbf{k}}\psi^{}_{\downarrow,\mathbf{k}})-{\Omega}(\psi^\dagger_{\uparrow,\mathbf{k}}\psi^{}_{\downarrow,\mathbf{k}}+\psi^{\dagger}_{\downarrow,\mathbf{k}}\psi^{}_{\uparrow,\mathbf{k}})-{\delta}(\psi^{\dagger}_{\uparrow,\mathbf{k}}\psi^{}_{\uparrow,\mathbf{k}}-\psi^{\dagger}_{\downarrow,\mathbf{k}}\psi^{}_{\downarrow,\mathbf{k}})\right];\\
	U&= \frac{1}{2V}\sum_{\mathbf{q,p,k}}\sum_{\sigma\sigma'}\left[g^{}_{\sigma\sigma'}\psi^{\dagger}_{\sigma,\mathbf{q+k}}\psi^{\dagger}_{\sigma',\mathbf{q-k}}\psi^{}_{\sigma',\mathbf{q+p}}\psi^{}_{\sigma,\mathbf{q-p}}\right]
\end{aligned}
\end{equation}
According to the standard Bogoliubov method, we have the quadratic Hamiltonian
\begin{equation}\label{before_transform}
	\frac{H}{V}=\frac{E_{\text{mf}}}{V}+\frac{1}{2V}\sum_{\sigma\sigma'}g^{2}_{\sigma\sigma'}n^{}_{\sigma} n^{}_{\sigma'}\sum_{\mathbf{k}\ne0}\frac{1}{2\epsilon^0_{\mathbf{k}}}-\frac{1}{2V}\sum_{\mathbf{k}\ne0}\Big(2\epsilon^0_\mathbf{k}+{\Omega}\frac{(n_{\uparrow}+n_{\downarrow})}{\sqrt{n_{\uparrow}n_{\downarrow}}}+g^{}_{\upuparrows}n^{}_{\uparrow}+g^{}_{\downdownarrows}n^{}_{\downarrow}\Big)+\frac{1}{2V}\sum_{\mathbf{k}\ne0}A^{\dagger}H_\text{Bog}A,
\end{equation}
where $A^{\dagger}=(\psi^{\dagger}_{\uparrow,\mathbf{k}},\psi^{}_{\uparrow,-\mathbf{k}},\psi^{\dagger}_{\downarrow,\mathbf{k}},\psi^{}_{\downarrow,-\mathbf{k}})$, and
\begin{equation}
	H_\text{Bog}=
    \begin{pmatrix}
		\epsilon^0_\mathbf{k}+{\Omega}\sqrt{\frac{n_{\downarrow}}{n_{\uparrow}}}+g_{\upuparrows}n_{\uparrow} & g_{\upuparrows}n_{\uparrow} & g_{\uparrow\!\downarrow}\sqrt{n_{\uparrow}n_{\downarrow}}-{\Omega} & g_{\uparrow\!\downarrow}\sqrt{n_{\uparrow}n_{\downarrow}} \\
		g_{\upuparrows}n_{\uparrow} & \epsilon^0_\mathbf{k}+{\Omega}\sqrt{\frac{n_{\downarrow}}{n_{\uparrow}}}+g_{\upuparrows}n_{\uparrow} & g_{\uparrow\!\downarrow}\sqrt{n_{\uparrow}n_{\downarrow}} & g_{\uparrow\!\downarrow}\sqrt{n_{\uparrow}n_{\downarrow}}-{\Omega} \\
		g_{\uparrow\!\downarrow}\sqrt{n_{\uparrow}n_{\downarrow}}-{\Omega} & g_{\uparrow\!\downarrow}\sqrt{n_{\uparrow}n_{\downarrow}} & \epsilon^0_\mathbf{k}+{\Omega}\sqrt{\frac{n_{\uparrow}}{n_{\downarrow}}}+g_{\downdownarrows}n_{\downarrow} & g_{\downdownarrows}n_{\downarrow} \\
		g_{\uparrow\!\downarrow}\sqrt{n_{\uparrow}n_{\downarrow}} & g_{\uparrow\!\downarrow}\sqrt{n_{\uparrow}n_{\downarrow}}-{\Omega} & g_{\downdownarrows}n_{\downarrow} & \epsilon^0_\mathbf{k}+{\Omega}\sqrt{\frac{n_{\uparrow}}{n_{\downarrow}}}+g_{\downdownarrows}n_{\downarrow}
	\end{pmatrix}.
\end{equation}
Under the Bogoliubov transformation, $H_\text{Bog}$ is diagonal in the basis of quasi-particle operators $\{b_{\pm,\mathbf{k}}\}$,
\begin{equation}\label{after_transform}
	\dfrac{1}{2}A^{\dagger}H_\text{Bog}A=\frac{1}{2}\mathcal{E}^{}_{+,\mathbf{k}}\Big(b^{\dagger}_{+,\mathbf{k}}b^{}_{+,\mathbf{k}}+b^{}_{+,-\mathbf{k}}b^{\dagger}_{+,-\mathbf{k}}\Big)+\frac{1}{2}\mathcal{E}^{}_{-,\mathbf{k}}\Big(b^{\dagger}_{-,\mathbf{k}}b^{}_{-,\mathbf{k}}+b^{}_{-,-\mathbf{k}}b^{\dagger}_{-,-\mathbf{k}}\Big).
\end{equation}
The Bogoliubov modes $\mathcal{E}_{\pm}$ satisfy
\begin{equation}\label{determinant}
	\bigg\|H_\text{Bog}-\mathcal{E}\begin{pmatrix}
		\sigma_z & 0\\
		0 & \sigma_z
	\end{pmatrix}\bigg\|=0.
\end{equation}
This gives
\begin{equation}\label{excitation}
	\mathcal{E}_{\{\pm\},\mathbf{k}}=\sqrt{D_\mathbf{k}\pm \sqrt{D_\mathbf{k}^2-\epsilon^0_\mathbf{k}\Big(\epsilon^0_\mathbf{k}+{\Omega}\frac{n_{\uparrow}+n_{\downarrow}}{\sqrt{n_{\uparrow}n_{\downarrow}}}\Big)\Big[\big(\epsilon^0_\mathbf{k}+2g_{\upuparrows}n_{\uparrow}+{\Omega}\sqrt{\frac{n_{\downarrow}}{n_{\uparrow}}}\big)\big(\epsilon^0_\mathbf{k}+2g_{\downdownarrows}n_{\downarrow}+{\Omega}\sqrt{\frac{n_{\uparrow}}{n_{\downarrow}}}\big)-\big(2g_{\uparrow\!\downarrow}\sqrt{n_{\uparrow}n_{\downarrow}}-{\Omega}\big)^2\Big]}}
\end{equation}
with
\begin{equation}
	D_\mathbf{k}=\frac{1}{2}\Big(\epsilon^0_\mathbf{k}+g_{\upuparrows}n_{\uparrow}+{\Omega}\sqrt{\frac{n_{\downarrow}}{n_{\uparrow}}}\Big)^2+\frac{1}{2}\Big(\epsilon^0_\mathbf{k}+g_{\downdownarrows}n_{\downarrow}+{\Omega}\sqrt{\frac{n_{\uparrow}}{n_{\downarrow}}}\Big)^2+\Big(g_{\uparrow\!\downarrow}\sqrt{n_{\uparrow}n_{\downarrow}}-{\Omega}\Big)^2-\frac{1}{2}\sum_{\sigma\sigma'}g_{\sigma\sigma'}^2 n_\sigma n_{\sigma'}.
\end{equation}
The total energy density that includes the quantum fluctuation can be obtained from Eq.\eqref{before_transform} and \eqref{after_transform} as
\begin{equation}\label{totalE}
	\frac{E}{V}=\frac{E_{\text{mf}}}{V}+\frac{1}{2V}\sum_{\mathbf{k}\ne0}\bigg(\sum_{\sigma\sigma'}g^{2}_{\sigma\sigma'}n^{}_{\sigma} n^{}_{\sigma'}\frac{1}{2\epsilon^0_{\mathbf{k}}}-\Big(2\epsilon^0_\mathbf{k}+{\Omega}\frac{(n_{\uparrow}+n_{\downarrow})}{\sqrt{n_{\uparrow}n_{\downarrow}}}+g^{}_{\upuparrows}n^{}_{\uparrow}+g^{}_{\downdownarrows}n^{}_{\downarrow}\Big)+\mathcal{E}_{+,\mathbf{k}} + \mathcal{E}_{-,\mathbf{k}}\bigg)
\end{equation}

In the large {\bf k}  limit, $\mathcal{E}_{\pm,\bf k}$ has the form $\sqrt{{\epsilon^0_{\bf k}}^2+c{\epsilon^0_{\bf k}}+c_0\pm \sqrt{c_1{\epsilon^0_{\bf k}}^2+c_2{\epsilon^0_{\bf k}+c_3}}}$, where $c, c_0, c_1, c_2$, and $c_3$ are lengthy expressions extracted from Eq.\eqref{excitation}, e.g., $c=g_{\upuparrows}n_\uparrow+g_{\downdownarrows}n_\downarrow+\Omega\frac{(n_\uparrow+n_\downarrow)}{\sqrt{n_\uparrow n_\downarrow}}$, $c_0=(g_{\upuparrows}+g_{\downdownarrows}-2g_{\uparrow\!\downarrow})\Omega\sqrt{n_\uparrow n_\downarrow} + \frac{(n_\uparrow+n_\downarrow)^2\Omega^2}{2n_\uparrow n_\downarrow}$, and $c_1=4c_0-c^2+2\sum_{\sigma\sigma'}g^2_{\sigma\sigma'}n_{\sigma}n_{\sigma'}$.
Then we have $\mathcal{E}_{+,\bf k} + \mathcal{E}_{-,\bf k} \sim 2\epsilon^0_{\bf k}+c+(c_0-\frac{c^2}{4}-\frac{c_1}{4})\frac{1}{\epsilon^0_{\bf k}} + O(\frac{1}{{\epsilon^0_{\bf k}}^2})$, and therefore the ultraviolet divergence for the integration of various terms in Eq.\eqref{totalE} can be exactly cancelled. 

Given $E(n_{\uparrow},n_{\downarrow})=E(n,S)$ (here $S=\cos\theta$ is the spin polarization), 
one can find the optimal spin polarization via ${\partial E(n,S)}/{\partial S}=0$.
In practice, we set $\delta g=0$ in numerically calculating quantum fluctuations in order to avoid the complex excitation spectra at small {\bf k}. In fact, for $\delta g<0$ the imaginary excitation spectra can be rectified by considering the high-order fluctuations, and such high-order terms  turn out to only produce little modification to the total fluctuation energy as long as $|\delta g|\ll |g_{ij}|$ [1].

\subsection{Effective two-body interaction at low-density limit}
In the low $n$ limit, the optimal spin polarization is given by the single-particle configuration, $\cos\theta=S_\text{sp}=\gamma$. 
Here we have the single-particle eigen-states: 
\begin{equation}
    \begin{pmatrix}
        \ket{-}\\\ket{+}
    \end{pmatrix}
    =
    \begin{pmatrix}
        \cos\frac{\theta}{2} & \sin\frac{\theta}{2}\\
        -\sin\frac{\theta}{2} & \cos\frac{\theta}{2}   
    \end{pmatrix}
    \begin{pmatrix}
        \ket{\uparrow}\\
        \ket{\downarrow}
    \end{pmatrix},
\end{equation}
and in such $\{+,-\}$ basis the original Hamiltonian \eqref{Hamiltonian-k-space} can be translated into
\begin{eqnarray}
	H&=&\sum_{\bf k}\left[\left(\epsilon^0_{\bf k}-\sqrt{\Omega^2+\delta^2}\right)\psi_{-,\bf k}^\dagger\psi_{-,\bf k} + \left(\epsilon^0_{\bf k}+\sqrt{\Omega^2+\delta^2}\right)\psi_{+,\bf k}^\dagger\psi_{+,\bf k}\right] + 
	\frac{1}{2V}\tilde{\sum}_{\bf \{k_i\}}\big\{\ g_1^{}\psi_{-,\bf k_1}^\dagger\psi_{-,\bf k_2}^\dagger\psi_{-,\bf k_3}\psi_{-,\bf k_4} \nonumber\\
	&& + g_2^{}\psi_{+,\bf k_1}^\dagger\psi_{+,\bf k_2}^\dagger\psi_{+,\bf k_3}\psi_{+,\bf k_4} + g_3^{}(\psi_{-,\bf k_1}^\dagger\psi_{+,\bf k_2}^\dagger\psi_{+,\bf k_3}\psi_{-,\bf k_4}+{\rm h.c}) + g_4^{}(\psi_{-,\bf k_1}^\dagger\psi_{+,\bf k_2}^\dagger\psi_{-,\bf k_3}\psi_{+,\bf k_4} + {\rm h.c}) \\
	&&+ g_5^{}(\psi_{-,\bf k_1}^\dagger\psi_{-,\bf k_2}^\dagger\psi_{+,\bf k_3}\psi_{+,\bf k_4} + {\rm h.c.}) + g_6^{}(\psi_{-,\bf k_1}^\dagger\psi_{-,\bf k_2}^\dagger\psi_{-,\bf k_3}\psi_{+,\bf k_4} + {\rm h.c.}) + g_7^{}(\psi_{+,\bf k_1}^\dagger\psi_{+,\bf k_2}^\dagger\psi_{+,\bf k_3}\psi_{-,\bf k_4} + {\rm h.c.})\big\}, \nonumber
\end{eqnarray}
where $\tilde{\sum}_{\bf \{k_i\}}$ includes momentum conservation.
The effective two-body interaction under the second-order perturbation is
\begin{equation}
g_\text{eff}^{(2)}
 =
\begin{tikzpicture}[line width=1pt,baseline=(v1.south)] 
    \begin{feynman} 
        \vertex (v1); 
        \vertex[right=1.3cm of v1] (v2); 
        \vertex at ($(v1)+(-0.4cm,+0.4cm)$) (i1); 
        \vertex at ($(v1)+(-0.4cm,-0.4cm)$) (i2); 
        \vertex at ($(v1)+(+0.4cm,+0.4cm)$) (f1); 
        \vertex at ($(v1)+(+0.4cm,-0.4cm)$) (f2); 
        \diagram* [small]{ 
            (i1)[label distance=-2pt] -- [dashed,edge label =\(-\),near start,inner sep = -0.5pt,shorten <=(0.0cm)] (v1) 
            -- [dashed, edge label=\(-\),shorten >=(0.0cm), near end,inner sep = -0.5pt] (f1), 
            (i2) -- [dashed,edge label'=\(-\),shorten <=(0.0cm),near start,inner sep = -0.5pt] (v1)  
            -- [dashed,edge label'=\(-\),shorten >=(0.0cm),near end,inner sep = -0.5pt] (f2), 
            }; 
        \draw[fill=black] (v1) circle (1.5pt); 
    \end{feynman} 
\end{tikzpicture}
+
\begin{tikzpicture}[line width=1pt,baseline=(v1.south)] 
    \begin{feynman} 
        \vertex (v1); 
        \vertex[right=1.3cm of v1] (v2); 
        \vertex at ($(v1)+(-0.4cm,+0.4cm)$) (i1); 
        \vertex at ($(v1)+(-0.4cm,-0.4cm)$) (i2); 
        \vertex at ($(v2)+(+0.4cm,+0.4cm)$) (f1); 
        \vertex at ($(v2)+(+0.4cm,-0.4cm)$) (f2); 
        \diagram* [small]{ 
            (i1)[label distance=-2pt] -- [dashed,edge label =\(-\),near start,inner sep = -0.5pt,shorten <=(0.0cm)] (v1) 
            -- [fermion,edge label=\(-\ \ \ \ -\),inner sep = 0.7pt,quarter left, looseness=1.2] (v2) -- [dashed, edge label=\(-\),shorten >=(0.0cm), near end,inner sep = -0.5pt] (f1), 
            (i2) -- [dashed,edge label'=\(-\),shorten <=(0.0cm),near start,inner sep = -0.5pt] (v1)  
            -- [fermion,edge label'=\(-\ \ \ \ -\), inner sep =0.7pt,quarter right, looseness=1.2] (v2) -- [dashed,edge label'=\(-\),shorten >=(0.0cm),near end,inner sep = -0.5pt] (f2), 
            }; 
        \draw[fill=black] (v1) circle (1.5pt); 
        \draw[fill=black] (v2) circle (1.5pt); 
    \end{feynman} 
\end{tikzpicture}
+
\begin{tikzpicture}[line width=1pt,baseline=(v1.south)] 
    \begin{feynman} 
        \vertex (v1); 
        \vertex[right=1.3cm of v1] (v2); 
        \vertex at ($(v1)+(-0.4cm,+0.4cm)$) (i1); 
        \vertex at ($(v1)+(-0.4cm,-0.4cm)$) (i2); 
        \vertex at ($(v2)+(+0.4cm,+0.4cm)$) (f1); 
        \vertex at ($(v2)+(+0.4cm,-0.4cm)$) (f2); 
        \diagram* [small]{ 
            (i1)[label distance=-2pt] -- [dashed,edge label =\(-\),near start,inner sep = -0.5pt,shorten <=(0.0cm)] (v1) 
            -- [fermion,edge label=\(+\ \ \ \ +\),inner sep = 0.7pt,quarter left, looseness=1.2] (v2) -- [dashed, edge label=\(-\),shorten >=(0.0cm), near end,inner sep = -0.5pt] (f1), 
            (i2) -- [dashed,edge label'=\(-\),shorten <=(0.0cm),near start,inner sep = -0.5pt] (v1)  
            -- [fermion,edge label'=\(+\ \ \ \ +\), inner sep =0.7pt,quarter right, looseness=1.2] (v2) -- [dashed,edge label'=\(-\),shorten >=(0.0cm),near end,inner sep = -0.5pt] (f2), 
            }; 
        \draw[fill=black] (v1) circle (1.5pt); 
        \draw[fill=black] (v2) circle (1.5pt); 
    \end{feynman} 
\end{tikzpicture}
+
\begin{tikzpicture}[line width=1pt,baseline=(v1.south)] 
    \begin{feynman} 
        \vertex (v1); 
        \vertex[right=1.3cm of v1] (v2); 
        \vertex at ($(v1)+(-0.4cm,+0.4cm)$) (i1); 
        \vertex at ($(v1)+(-0.4cm,-0.4cm)$) (i2); 
        \vertex at ($(v2)+(+0.4cm,+0.4cm)$) (f1); 
        \vertex at ($(v2)+(+0.4cm,-0.4cm)$) (f2); 
        \diagram* [small]{ 
            (i1)[label distance=-2pt] -- [dashed,edge label =\(-\),near start,inner sep = -0.5pt,shorten <=(0.0cm)] (v1) 
            -- [fermion,edge label=\(+\ \ \ \ +\),inner sep = 0.7pt,quarter left, looseness=1.2] (v2) -- [dashed, edge label=\(-\),shorten >=(0.0cm), near end,inner sep = -0.5pt] (f1), 
            (i2) -- [dashed,edge label'=\(-\),shorten <=(0.0cm),near start,inner sep = -0.5pt] (v1)  
            -- [fermion,edge label'=\(-\ \ \ \ -\), inner sep =0.7pt,quarter right, looseness=1.2] (v2) -- [dashed,edge label'=\(-\),shorten >=(0.0cm),near end,inner sep = -0.5pt] (f2), 
            }; 
        \draw[fill=black] (v1) circle (1.5pt); 
        \draw[fill=black] (v2) circle (1.5pt); 
    \end{feynman} 
\end{tikzpicture},
\end{equation}
where the vertices are
$
		\begin{tikzpicture}[line width=1pt,baseline=(v1.south)] 
			\begin{feynman} 
				\vertex (v1); 
				\vertex[right=1.3cm of v1] (v2); 
				\vertex at ($(v1)+(-0.4cm,+0.4cm)$) (i1); 
				\vertex at ($(v1)+(-0.4cm,-0.4cm)$) (i2); 
				\vertex at ($(v1)+(+0.4cm,+0.4cm)$) (f1); 
				\vertex at ($(v1)+(+0.4cm,-0.4cm)$) (f2); 
				\diagram* [small]{ 
					(i1)[label distance=-2pt] -- [dashed,edge label =\(-\),near start,inner sep = -0.5pt,shorten <=(0.0cm)] (v1) 
					-- [dashed, edge label=\(-\),shorten >=(0.0cm), near end,inner sep = -0.5pt] (f1), 
					(i2) -- [dashed,edge label'=\(-\),shorten <=(0.0cm),near start,inner sep = -0.5pt] (v1)  
					-- [dashed,edge label'=\(-\),shorten >=(0.0cm),near end,inner sep = -0.5pt] (f2), 
					}; 
				\draw[fill=black] (v1) circle (1.5pt); 
			\end{feynman} 
		\end{tikzpicture}
		\equiv g_1 = g_{\text{eff,mf}},\quad
		\begin{tikzpicture}[line width=1pt,baseline=(v1.south)] 
			\begin{feynman} 
				\vertex (v1); 
				\vertex[right=1.3cm of v1] (v2); 
				\vertex at ($(v1)+(-0.4cm,+0.4cm)$) (i1); 
				\vertex at ($(v1)+(-0.4cm,-0.4cm)$) (i2); 
				\vertex at ($(v1)+(+0.4cm,+0.4cm)$) (f1); 
				\vertex at ($(v1)+(+0.4cm,-0.4cm)$) (f2); 
				\diagram* [small]{ 
					(i1)[label distance=-2pt] -- [dashed,edge label =\(-\),near start,inner sep = -0.5pt,shorten <=(0.0cm)] (v1) 
					-- [dashed, edge label=\(+\),shorten >=(0.0cm), near end,inner sep = -0.5pt] (f1), 
					(i2) -- [dashed,edge label'=\(-\),shorten <=(0.0cm),near start,inner sep = -0.5pt] (v1)  
					-- [dashed,edge label'=\(+\),shorten >=(0.0cm),near end,inner sep = -0.5pt] (f2), 
					}; 
				\draw[fill=black] (v1) circle (1.5pt); 
			\end{feynman} 
		\end{tikzpicture}
		\equiv g_5= g_0\sin^2 {\theta}, \quad \text{and} \quad
		\begin{tikzpicture}[line width=1pt,baseline=(v1.south)] 
			\begin{feynman} 
				\vertex (v1); 
				\vertex[right=1.3cm of v1] (v2); 
				\vertex at ($(v1)+(-0.4cm,+0.4cm)$) (i1); 
				\vertex at ($(v1)+(-0.4cm,-0.4cm)$) (i2); 
				\vertex at ($(v1)+(+0.4cm,+0.4cm)$) (f1); 
				\vertex at ($(v1)+(+0.4cm,-0.4cm)$) (f2); 
				\diagram* [small]{ 
					(i1)[label distance=-2pt] -- [dashed,edge label =\(-\),near start,inner sep = -0.5pt,shorten <=(0.0cm)] (v1) 
					-- [dashed, edge label=\(-\),shorten >=(0.0cm), near end,inner sep = -0.5pt] (f1), 
					(i2) -- [dashed,edge label'=\(-\),shorten <=(0.0cm),near start,inner sep = -0.5pt] (v1)  
					-- [dashed,edge label'=\(+\),shorten >=(0.0cm),near end,inner sep = -0.5pt] (f2), 
					}; 
				\draw[fill=black] (v1) circle (1.5pt); 
			\end{feynman} 
		\end{tikzpicture}
		\equiv {g_6}/{\sqrt{2}}=\sqrt{2}\sin\!\theta(\cos\!\theta-\beta)g_0.
$
The internal lines refer to the single-particle Green's functions: 
\begin{equation}
		G^0_{--}(p) = {1}/{(p^0-\epsilon_p^0 +i0^+)}\ \ \text{and} \ \ \ 
		G^0_{++}(p) = {1}/{(p^0-\epsilon_p^0-{2\sqrt{\Omega^2+\delta^2}} + i 0^+)}.
\end{equation}
With the renormalization relations $g_{\sigma\sigma'}\to g_{\sigma\sigma'}+g_{\sigma\sigma'}^2\int \frac{1}{2\epsilon^0_k} \frac{d^3k}{(2\pi)^3}$ and the fact $\sum_{\sigma\sigma'}g_{\sigma\sigma'}^2n_{\sigma}n_{\sigma'}/n^2=g_1^2 + g_5^2 + g_6^2/2$,
we can straightforwardly perform
\begin{equation}
	\begin{aligned}
		g_\text{eff}^{(2)}
		&= g_\text{eff,mf} + g_1^2\int\!\frac{d^3k}{(2\pi)^3}\left(i\!\int\! \frac{dk^0}{2\pi}G^0_{--}(k)G^0_{--}(-k)+\frac{1}{2\epsilon^0_k}\right)
		\\
		&\ \ \ + 
		g_5^2\int\!\frac{d^3k}{(2\pi)^3}\left(i\!\int\! \frac{dk^0}{2\pi}G^0_{++}(k)G^0_{++}(-k)+\frac{1}{2\epsilon^0_k}\right) + 
		\frac{g_6^2}{2}\int\!\frac{d^3k}{(2\pi)^3}\left(i\!\int\! \frac{dk^0}{2\pi}G^0_{++}(k)G^0_{--}(-k)+\frac{1}{2\epsilon^0_k}\right),
	\end{aligned}
\end{equation}
and obtain the effective interaction
\begin{equation}
	g_\text{eff}^{(2)}=g_\text{eff,mf}+\frac{(1-\gamma^2)^{3/4}}{2\pi}\left[1-\gamma^2+\sqrt{2}(\gamma-\beta)^2\right]\eta^{1/2}g_0.
\end{equation}
Note that $g_\text{eff}^{(2)}$ above is exactly the effective interaction for the lowest dressed branch of the single-particle states. It is composed by two parts: $g_\text{eff,mf}$ is the mean-field contribution, and the second part is  contributed from the quantum fluctuations (or Lee-Huang-Yang(LHY) corrections), as shown by the red dot line in Fig.1(c) and Fig.\ref{figS1} at $n\rightarrow 0$. In the special case $\beta=\gamma$, this part reproduces the LHY-induced two-body interaction as discussed in Ref.[2].
Here the existence of (meta)stable gas requires $g_\text{eff}^{(2)}>0$, and the boundary $g_\text{eff}^{(2)}=0$ is shown as the black dashed line in Fig.2 (separating regions I and II).

\subsection{Spin twist and double minima structure}

\begin{figure}[!ht]
    \centering
    \includegraphics[width=0.8\linewidth]{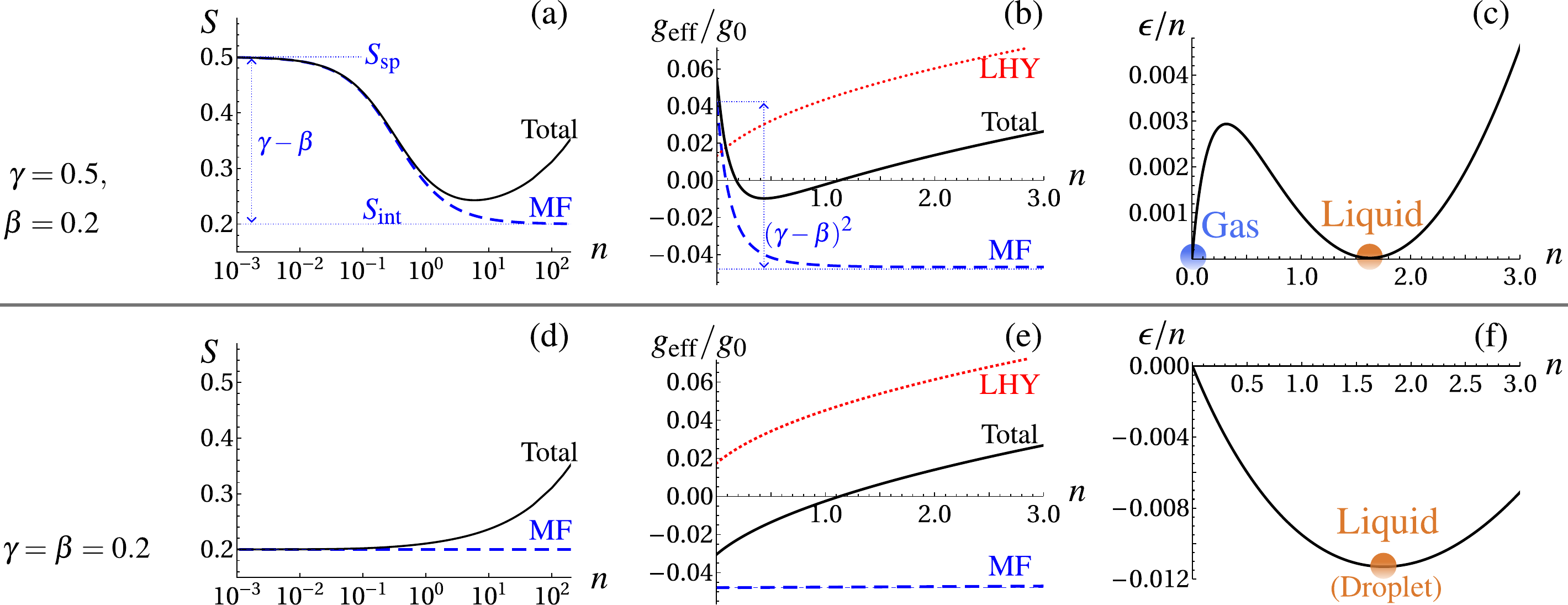}
    \caption{The polarization, effective interaction and energy per particle curves, with (upper panel) and without (lower panel) spin-twist. The case of spin twist is identical to that shown in Fig.1(b,c,d).  Without spin-twist, the total effective interaction $g_{\rm eff}$ always increases monotonically, and the energy per particle $\epsilon/n$ presents only one minimum, namely gas or droplet state. $\alpha$ and $\eta$ are the same as in Fig.1.}
    \label{figS1}
\end{figure}

To demonstrate the vital role of spin twist played in the liquid-gas coexistence, in Fig.\ref{figS1} we compare the $\epsilon/n (n)$ curves  with and without spin twist.  Fig.\ref{figS1}(a,b,c) are for the case with spin twist ($\beta\ne\gamma$), which are identical to Fig.1(b,c,d) in the main text. It shows that the spin twist can result in an additional mean-field repulsion in the low-$n$ regime, which uniquely stabilizes the gas state and facilitates  the liquid-gas coexistence in view of the double minima structure of $\epsilon/n (n)$ curve. In comparison,  Fig.\ref{figS1}(d,e,f) are for the case without spin twist ($\beta=\gamma$), where the mean-field contributions to $S$ and $g_{\rm eff}$ are both static for all densities and the resulted $\epsilon/n (n)$ can only show one minimum (representing either liquid or gas) but not two. In the latter case, the liquid and gas cannot be (locally) stable simultaneously and thus cannot coexist.

\section{Liquid-gas coexistence in a harmonic trap}

Here we present more details on the coexistence of liquid and gas in an isotropic harmonic trap $V({\cp r})$, as shown by Fig.3 in the main text. 

\begin{figure}[!ht]
    \centering
 \includegraphics[width=17cm]{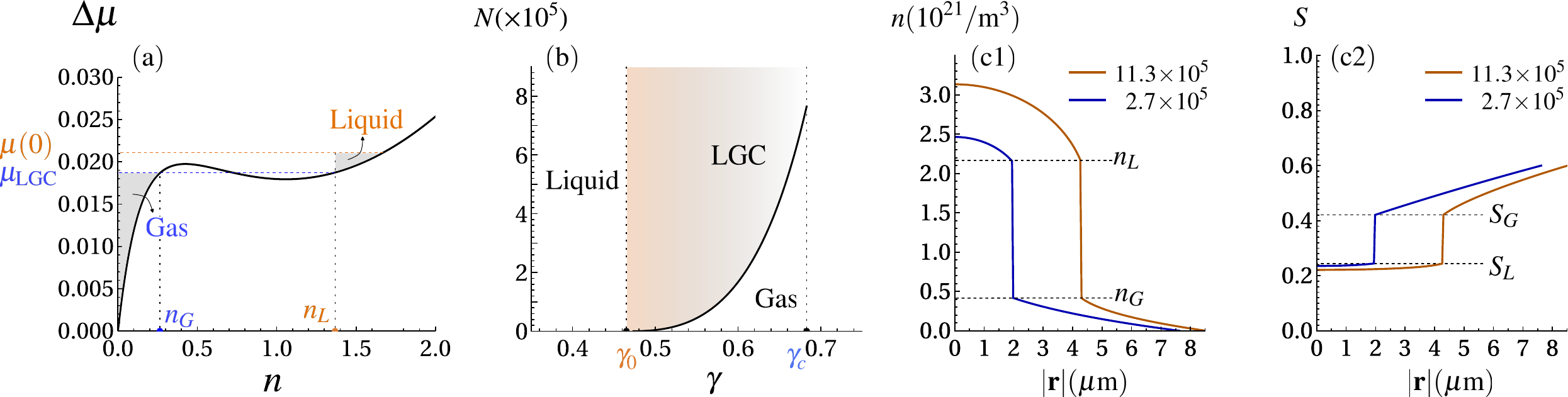}
\caption{Liquid-gas coexistence at fixed $\alpha=-0.11,\beta=0.141,\eta=0.0157$.  (a) Shifted chemical potential $\Delta\mu\equiv \mu+\sqrt{\Omega^2+\delta^2}$ as a function of $n$ at $\gamma=0.6$. The blue horizontal line marks the shifted $\mu_{\rm LGC}$ at liquid-gas coexistence, and the orange horizontal line marks the shifted chemical potential $\mu(0)$ at trap center. The gas (liquid) state populates $\mu<\mu_{\rm LGC}$ ($\mu>\mu_{\rm LGC}$). (b) Phases tuned by $\gamma$ and boson number $N$ in an isotropic harmonic trap with frequency $\omega=(2\pi)50$Hz. The coexistence occurs in the colored region with $N>N_c$ and $\mu(0)>\mu_{\rm LGC}$. (c1,c2): Profiles of total density $n$ and spin polarization $S$ for different boson numbers $N$ at fixed $\gamma=0.6$. At the liquid-gas interface $n_{L,G}$ and $S_{L,G}$ are universal values for all $N(>N_c)$. For all plots, we scale the density and energy per particle respectively by $\Omega/g_0$ and $\Omega$.}
    \label{supple_LGC}
\end{figure}

At the coexistence of liquid and gas, let us denote $\mu(n_L)=\mu(n_G)\equiv \mu_{\rm LGC}$. Then according to the $P (1/n)$ and $\mu (n)$ curves in Fig.3(a,b), we can see that the gas phase stays for $\mu<\mu_{\rm LGC}$, as it has a higher pressure (and thus a lower thermodynamic potential) than liquid; while the liquid phase stays for $\mu>\mu_{\rm LGC}$, at it has a higher pressure than gas.  In Fig.\ref{supple_LGC}(a), we show the occupation of gas and liquid  in the $\mu$-$n$ plane taking a typical value of $\gamma\in(\gamma_0,\gamma_c)$. 

Under the local density approximation $\mu({\cp r})=\mu(0)-V({\cp r})$, the local $\mu$ decays from the trap center to edge. Then it follows that the liquid phase with higher $\mu(>\mu_{\rm LGC})$ locates at the trap center, and the gas phase with lower $\mu(<\mu_{\rm LGC})$ locates at the edge. To enable the coexistence of liquid and gas in the trap, the chemical potential $\mu(0)$ at the trap center should be higher than $\mu_{\rm LGC}$, which requires the boson number $N$ be larger than a critical value $N_c$. In other words, at $N=N_c$, $\mu(0)$ reaches $\mu_{\rm LGC}$ and a liquid phase starts to emerge at the trap center. In Fig.\ref{supple_LGC}(b), we plot out $N_c$ as a function of $\gamma$ (solid line). The  liquid-gas coexistence  occurs within $\gamma\in(\gamma_0,\gamma_c)$ once  $N>N_c$, as shown by the colored region therein.    

As shown by Fig.3(d) in the main text,  at the liquid-gas interface the total density $n$ undergoes a discontinuous jump from $n_L$ to $n_G$, and the spin polarization $S$ jumps from $S_L$ to $S_G$. Here we emphasize that such discontinuities are universal for different $N (>N_c)$. As shown by Fig.\ref{supple_LGC}(c1,c2), different $N$ would not alter $n_{L,G}$ and $S_{L,G}$ at the liquid-gas interface, but just change the size of liquid phase in the trap. Therefore $n_{L,G}$ and $S_{L,G}$ can serve as universal quantities to characterize the liquid-gas  coexistence for thermodynamic systems.


\section{Critical exponents near the melting of liquid-gas coexistence}

Inspired by the classical treatment of temperature-driven liquid-gas coexistence in Ref.[3],
here we introduce the mixed thermodynamic function $W(x,\mu,n) = \epsilon(x,\mu,n) -\mu n$, which determines the equilibrium state via $\partial W/\partial n=0$ and the coexistence critical point via $\partial^2W(x_c,\mu_c,n_c)/\partial n^2=\partial^3W(x_c,\mu_c,n_c)/\partial n^3=0$. The variable $x$ here can represent any tunable parameter we are interested in, such as $\gamma$ or $\alpha$ in this work. 

We expand $W(x,\mu,n)$ near the critical point to the fourth-order of $(n-n_c)$:
\begin{equation}\label{n-expansion}
    W(x,\mu,n)= W(x,\mu,n_c)+ \sum_{i=1}^{i=4}w_i(n-n_c)^i,
\end{equation}
where $w_4(x,\mu,n_c)$ is finite, and $w_{1,2,3}(x,\mu,n_c)$ is close to zero and can be expanded in terms of $(x-x_c)$ and $(\mu-\mu_c)$,
\begin{equation}
		w_1 = \frac{\partial W(x,\mu,n_c)}{\partial n}= \frac{\partial^2 \epsilon(x_c,n_c)}{\partial n \partial x}(x-x_c) - (\mu - \mu_c),\quad
		w_{\{i=2,3\}}^{} = \frac{\partial^i W(x,\mu,n_c)}{i!\ \partial n^i} =\frac{\partial^{i+1} \epsilon(x_c,n_c)}{i!\ \partial n^{i} \partial x}(x-x_c).
\end{equation}
Denoting $\phi = (n_L-n_G)/2$ and $\phi_0 = (n_L+n_G)/2-n_c$, we have
\begin{eqnarray}\label{replacement}
	n_L-n_c = \phi_0+\phi \quad \text{and} \quad n_G-n_c=\phi_0-\phi.
\end{eqnarray}
Substituting Eq.\eqref{replacement} into Eq.\eqref{n-expansion}, and rearranging the expansion in terms of $\phi^i$, we have 
\begin{equation}\label{phi-expansion}
\begin{aligned}
	W(x,\mu,n_L) &= W(x,\mu,n_c)+\sum_{i=1}^{i=4}\left(w_i\phi_0^i + \tilde{w}_i\phi^i\right),\\
	W(x,\mu,n_G) &= W(x,\mu,n_c)+\sum_{i=1}^{i=4}\left(w_i\phi_0^i + (-1)^i\tilde{w}_i\phi^i\right),
\end{aligned}
\end{equation}
where the coefficients are
\begin{equation}\label{coefficients}
    \begin{aligned}
        \tilde{w}_1&= w_1+2w_2\phi_0+3w_3\phi_0^2+4w_4\phi_0^3,\\
        \tilde{w}_2&=w_2+3w_3\phi_0+6w_4\phi_0^2,\\
        \tilde{w}_3&=w_3+4w_4\phi_0,\\
		\tilde{w}_4&=w_4.
    \end{aligned}
\end{equation}
At coexistence, $n_L$ and $n_G$ share the same pressure $P(n_L)=P(n_G)\equiv-W(x,\mu,n)_{\rm min}$. This mean that for Eq.\eqref{phi-expansion} with a given $\phi_0$, we have $W(x,\mu,n_L) = W(x,\mu, n_G)$.
Therefore, the terms of the odd-order of $\phi$ are absent in Eq.\eqref{phi-expansion}, i.e., $\tilde{w}_1=\tilde{w}_3=0$, which leads to
\begin{equation}\label{averaged_n}
	\phi_0 = \frac{-w_3}{4w_4}=\frac{-1}{4w_4}\frac{\partial^4 \epsilon(x_c,n_c)}{\partial n^3\partial x}(x-x_c).
\end{equation}
The function $W(x,\mu,n)$ in Eq.\eqref{phi-expansion} is then reduced to $W_0(x,\mu,\phi_0)+\tilde{w}_2\phi^2+w_4\phi^4$, whose minimum describes the equilibrium state with 
\begin{equation}\label{relative_n}
        \phi^2 = \frac{-\tilde{w}_2}{2w_4} = \frac{-1}{2w_4}\frac{\partial^3 \epsilon(x_c,n_c)}{\partial n^2\partial x}(x-x_c)+O\left((x-x_c)^2\right).
\end{equation}
Eq.\eqref{averaged_n} and \eqref{relative_n} give the critical exponents $\xi=1$ and $\lambda=1/2$. 

\begin{figure}[!ht]
    \centering
    \includegraphics[width=0.55\linewidth]{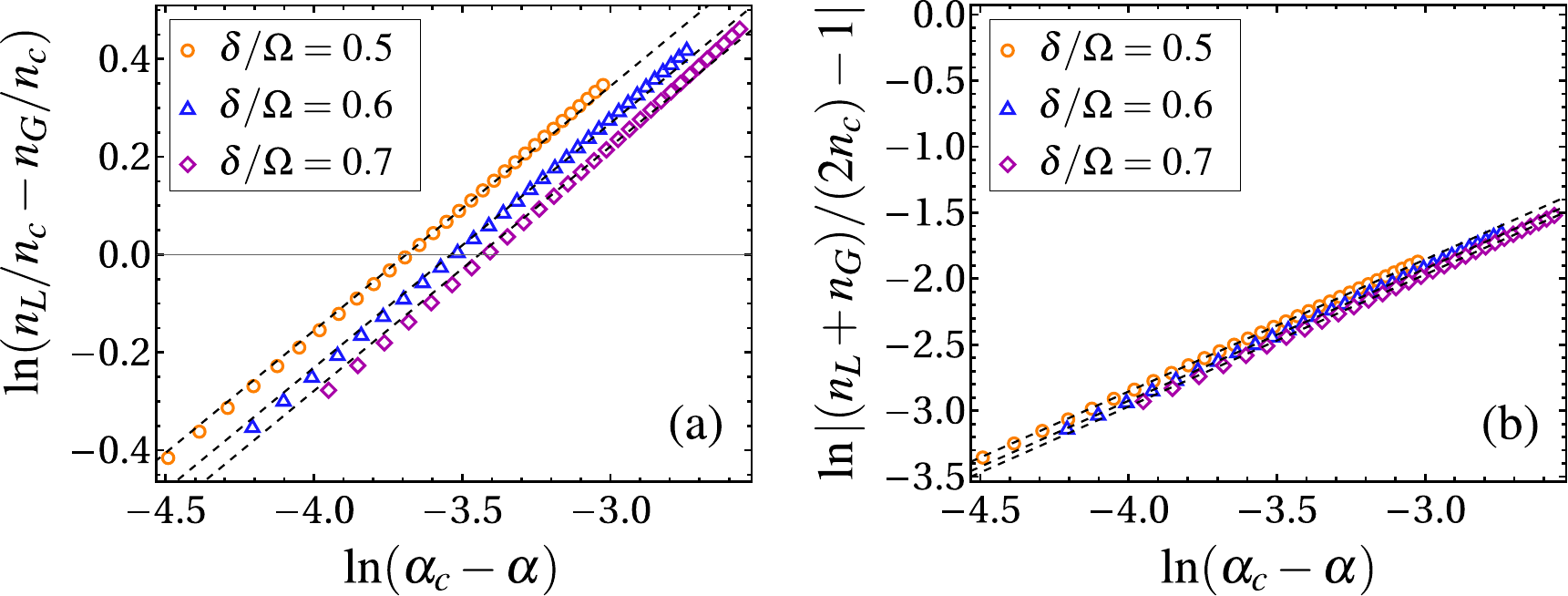}
    \caption{Universal critical scaling for the relative (a) and averaged (b) densities of liquid and gas at their coexistence. Here we consider the liquid-gas coexistence tuned by $\alpha$. The slopes of all fitting lines in  the (a) are $1/2$ and in (b) are $1$, giving the universal exponents $\lambda=1/2$ and $\xi=1$. $\beta,\eta$ are the same as in Fig.4.}
    \label{figS2}
\end{figure}

Note that above derivation applies to any tunable parameter $x$. In the main text we have chosen $x=\gamma$, while keeping other parameters fixed. In Fig.\ref{figS2}, we choose $x=\alpha$ as the tunable parameter, and find that the two critical exponents $\xi=1$ and $\lambda=1/2$ still apply in this case. Therefore, these exponents are universal ones in characterizing the liquid-gas coexistence in our system. 

\begin{@fileswfalse}

\end{@fileswfalse}

\end{document}